\documentclass[twocolumn,showpacs,nopreprintnumbers,amsmath,pra]{revtex4}
\usepackage{graphicx}
\usepackage{amssymb}

\begin{document}
\title{A high-sensitivity laser-pumped $M_x$-magnetometer}
\author{S. Groeger, J.-L. Schenker, R. Wynands\footnote{Present
address: PTB 4.41, 38116 Braunschweig, Germany}, A. Weis}
\affiliation{Universit\'e de Fribourg, Chemin de Mus\'ee 3, 1700
Fribourg, Switzerland}
\date{\today}

\begin{abstract}

We discuss the design and performance of a laser-pumped cesium vapor
magnetometer in the $M_x$-configuration. The device will be employed
in the control and stabilization of fluctuating magnetic fields and
gradients in a new experiment searching for a permanent electric
dipole moment of the neutron. We have determined the intrinsic
sensitivity of the device to be 29\,fT in a 1\,Hz bandwidth, limited
by low frequency noise of the laser power. In the shot noise limit
the magnetometer can reach a sensitivity of 10\,fT in a 1\,Hz
bandwidth. We have used the device to study the fluctuations of a
stable magnetic field in a multi-layer magnetic shield for
integration times in the range of 2-100 seconds. The residual
fluctuations for times up to a few minutes are traced back to the
instability of the power supply used to generate the field.
\end{abstract}

\pacs{07.55.Ge, 32.30.Dx, 32.70.Jz, 33.55.-b} \maketitle

\section{Introduction}
\label{sec:intro}

In many areas of fundamental and applied science the sensitive
detection of weak magnetic fields and small field fluctuations is of
great importance. In the applied sector this concerns, for instance,
non-destructive testing of materials \cite{KFA}, geomagnetic and
archaeological prospecting \cite{becker}, and the expanding field of
biomagnetism \cite{andrae}. In the realm of fundamental physics,
strong demands on magnetometric sensitivity are placed by modern
experiments looking for small violations of discrete symmetries in
atoms and elementary particles. For instance, many experiments
searching for time-reversal or parity violation rely on the precise
monitoring and control of magnetic fields, with the sensitivity of
the overall experiment directly related to the ultimate sensitivity
and stability of the magnetic field detection. Picotesla or even
femtotesla sensitivity requirements for averaging times of seconds to
minutes are common in that field.

Our particular interest in this respect lies in the search for a
permanent electric dipole moment (EDM) of the neutron. Such a moment
violates both time reversal invariance and parity conservation. A
finite sized EDM would seriously restrict theoretical models that
extend beyond the standard model of particle physics \cite{Alt96}.
Recently our team has joined a collaboration aiming at a new
measurement of the permanent EDM of ultra-cold neutrons (UCN) to be
produced from the UCN source under construction at
Paul-Scherrer-Institute in Switzerland \cite{SUNS}. A neutron EDM
spec\-tro\-me\-ter will be used, in which the neutron spin-flip
frequency will be measured by a Ramsey resonance method in UCN
storage chambers exposed to a homogenous magnetic field. Each neutron
chamber has two compartments in which the neutrons are exposed to a
static electric field oriented parallel/antiparallel to the magnetic
field. The signature of a finite EDM will be a change of the neutron
Larmor frequency that is synchronous with the reversal of the
relative orientations of the magnetic and electric fields. Magnetic
field instabilities and inhomogeneities may mimic the existence of a
finite neutron EDM. The control of such systematic effects is
therefore a crucial feature of the EDM experiment. It is planned to
use a set of optically pumped cesium vapor magnetometers (OPM),
operated in the $M_x$ configuration \cite{Blo62,aleMx} to perform
that control.

Although OPMs pumped by spectral discharge lamps are suited for the
task, we have opted for a system of laser pumped OPMs (LsOPM). It was
shown previously that the replacement of the lamp in an OPM by a
resonant laser can lead to an appreciable gain in magnetometric
sensitivity \cite{aleMx,ale2}. Laser pumping further offers the
advantage that a single light source can be used for the simultaneous
operation of several dozens of magnetometer heads. In that spirit we
have designed and tested a LsOPM with a geometry compatible with the
neutron EDM experiment. In this report we present the design and the
performance of the Cs-LsOPM operated in a phase-stabilized mode and
discuss a systematic effect specifically related to laser pumping.

\section{The optically-pumped $M_x$ magnetometer}

Optically pumped magnetometers can reach extreme sensitivities of a
few fT$/\sqrt\mathrm{Hz}$ \cite{aleMx}, comparable to standard SQUID
(superconducting quantum interference device) detectors. Recently a
low field OPM with a sub-fT resolution was demonstrated
\cite{romalis_Nature}. The use of OPMs for the detection of
biomagnetic signals was recently demonstrated by our group
\cite{georg,georg2}.

As a general rule the optimum choice of the OPM depends on the
specific demands (sensitivity, accuracy, stability, bandwidth,
spatial resolution, dynamic range, etc.) of the magnetometric problem
under consideration. In our particular case the main requirements are
a highest possible sensitivity and stability for averaging times
ranging from seconds up to 1000 seconds in a $2\,\mu\mbox{T}$ field
together with geometrical constraints imposed by the neutron EDM
experiment.

Optically pumped alkali vapor magnetometers rely on an optical-radio
frequency (r.f.) resonance technique and are described, e.g., in
\cite{Blo62}. When an alkali vapor is irradiated with circularly
polarized light resonant with the D$_1$ absorption line (transition
from the nS$_{1/2}$ ground state to the first nP$_{1/2}$ excited
state), the sample is optically pumped and becomes spin polarized
(magnetized) along the direction of the pumping light. While lamp
pumped OPMs simultaneously pump all hyperfine transitions of the
D$_1$ line, the use of a monomode laser in a LsOPM allows one to
resolve the individual hyperfine transitions provided that their
Doppler width does not exceed the hyperfine splitting in both the
excited and the ground states. This is, for example, the case for
the D$_1$ transition of the alkali isotopes $^{133}$Cs and
$^{87}$Rb. In that case it is advantageous to set the laser
frequency to the $F=I+1/2\,\rightarrow\,F=I-1/2$ transition, which
allows one to optically pump the atoms into the two (non-absorbing)
dark states $|nS_{1/2}; F; M_F=F,F-1\rangle$ using $\sigma^{+}$
polarized radiation. A magnetic field $\mathbf{B}_1(t)$ oscillating
at the frequency $\omega_{\mathrm{rf}}$, which is resonant with the
Zeeman splitting of the states, drives population out of the dark
states into absorbing states, so that the magnetic resonance
transition can be detected via a change of the optical transmission
of the vapor. That is the very essence of optically detected
magnetic resonance.

\begin{figure}
\includegraphics{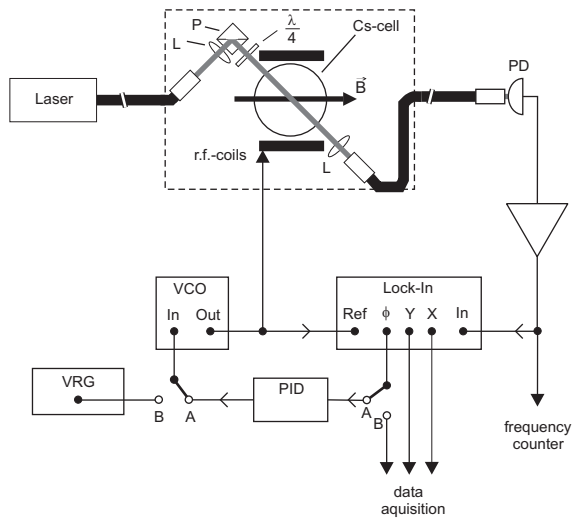}%[width=7cm,keepaspectratio]
\caption{Schematic setup of the phase-stabilized magnetometer in the
closed-loop (A) and the scanning (B) mode. The dashed box indicates
the sensor head. L: lens, P: polarizing beamsplitter, $\lambda$/4:
quarter-wave plate, PD: photodiode, VRG: voltage ramp generator, VCO:
voltage-controlled oscillator, PID: feedback amplifier. The
stabilization system of the laser frequency is not shown.}
\label{fig:soplscheme}
\end{figure}

In the so-called $M_x$ or $45^\circ$ configuration the static
magnetic field $B_0$ to be measured is oriented at 45$^\circ$ with
respect to the laser beam, while the oscillating magnetic field
$\mathbf{B}_1(t)$ is at right angles with respect to $\mathbf{B}_0$
(Fig.~\ref{fig:soplscheme}). In classical terms, the Larmor
precession of the magnetization around $\mathbf{B}_0$ (at the
frequency $\omega_L$) is driven by the co-rotating component of the
$\mathbf{B}_1(t)$-field, which imposes a phase on the precessing
spins. The projection of the precessing polarization onto the
propagation direction of the light beam then leads to an oscillating
magnetization component along that axis, and therefore to a periodic
modulation of the optical absorption coefficient. The system behaves
like a classical oscillator, in which the amplitude and the phase of
the response (current from a photodiode detecting the transmitted
laser intensity) depend in a resonant way on the frequency of the
$B_1$ field. From the resonance condition
$\omega_L=\omega_{\mathrm{rf}}$ the Larmor frequency and hence the
magnetic field can be inferred.

When the AC component of the detected optical signal is transmitted
to the coils producing the $\mathbf{B}_1(t)$ field with a 90$^\circ$
phase shift and an appropriate gain, the system will spontaneously
oscillate at the resonance frequency. In that self-oscillating
configuration the OPM can in principle follow changes of the magnetic
field instantaneously with a bandwidth limited by the Larmor
frequency only \cite{Blo62}.

Here we have used an alternative mode of operation, the so-called
phase-stabilized mode. The in-phase amplitude $X$, the quadrature
amplitude $Y$ and the phase $\phi$ of the photocurrent with respect
to the oscillating magnetic field are given by

\begin{eqnarray}
    \mathrm{a)}~~~& X(x)~=&  -A\, \frac{x}{x^2+1+S}\\
    \mathrm{b)}~~~& Y(x)~=&  -A\, \frac{1}{x^2+1+S}\\
    \mathrm{c)}~~~& \phi(x)~=& \arctan x\,,
    \label{eq:resonance}
\end{eqnarray}
where $x=(\nu_L-\nu_{\mathrm{rf}})/\Delta \nu_{\mathrm{HWHM}}$ is
the detuning normalized to the (light-power dependent) half width at
half maximum $\Delta \nu_{\mathrm{HWHM}}$ of the resonance. $S$ is a
saturation parameter which describes the r.f.~power broadening of
the line. It is interesting to note that the width of the phase
dependence, which is determined by the ratio of the $X(x)$ and
$Y(x)$ signals, is independent of $S$, and hence immune to
r.f.~power broadening. The phase $\phi(x)$ changes from $0^\circ$ to
$-180^\circ$ as $\nu_{\mathrm{rf}}$ is tuned over the Larmor
frequency. Near resonance the phase is $-90^\circ$ and has a linear
dependence on the detuning $\nu_L-\nu_{\mathrm{rf}}$. $\phi(x)$ is
detected by a phase sensitive amplifier (lock-in detector) whose
phase output drives a voltage-controlled oscillator (VCO) which
feeds the r.f.~coils. The VCO signal, phase shifted by $90^\circ$,
serves as a reference to the phase detector. This feedback loop thus
actively locks the r.f.~frequency to the Larmor frequency and the
magnetometer tracks magnetic field changes in a phase coherent
manner. That mode of operation is a modification of the
self-oscillating magnetometer in the sense that the lock-in
amplifier, the loop filter (PID), and the VCO represent the
components of a tracking filter which shifts the detected signal by
$90^\circ$ and applies the filtered signal to the r.f. coils. The
differences to the self-oscillating scheme are the following: the
bandwidth of the phase-stabilized magnetometer is determined by the
transmission function of the feedback loop, and the phase shift is
always $90^\circ$ independent of the Larmor frequency, while in the
self-oscillating scheme the phase-shifter has a frequency dependence
and is $90^\circ$ only for a single Larmor frequency. Note that the
tracking filter in a strict sense is not a phase-locked loop (PLL),
since there is only one detectable frequency in the system, i.e.,
$\nu_\mathrm{rf}$. A detuning between the r.f. frequency and the
Larmor frequency produces a static phase shift, while in a real PLL
the detuning between the reference frequency and the frequency which
is locked produces a time dependent phase shift.

\section{Magnetometer hardware}
\label{sec:setup}

The LsOPM for the n-EDM experiment consists of two parts: a sensor
head containing no metallic parts except the r.f.~coils, and a base
station mounted in a portable 19'' rack drawer, which contains the
frequency stabilized laser and the photodetector. The laser light is
carried from the base station to the sensor head by a 10\,m long
multimode fiber with a core diameter of $800\,\mu$m. The light
transmitted through the cell is carried back to the detection unit
by a similar fiber. The sensor head is designed to fit into a tube
of 104\,mm diameter, coaxial with the $2\,\mu$T field, and has a
total length of 242\,mm. The main component of the sensor is an
evacuated glass cell with a diameter of 7\,cm containing a droplet
of cesium in a sidearm connected to the main volume. A constriction
in the sidearm minimizes the collision rate of vapor atoms with the
cesium metal. The probability of spin depolarization due to wall
collisions with the inner surface of the glass cell is strongly
reduced by a thin layer of paraffin coating the cell walls. The cell
was purchased from MAGTECH Ltd., St. Petersburg, Russia. A pair of
circular coils (70\,mm diameter separated by 52\,mm) encloses the
cell and produces the oscillating magnetic field $\mathbf{B}_1(t)$.

The light driving the magnetometer is produced by a tunable
extended-cavity diode laser in Littman configuration (Sacher
Lasertechnik GmbH, model TEC500). The laser frequency is actively
locked to the 4-3 hyperfine component of the Cs D$_1$ transition
($\lambda = 894\,\mbox{nm}$) in an auxiliary cesium vapor cell by
means of the dichroic atomic vapor laser lock (DAVLL) technique
\cite{DAVLL}. The stabilization to a Doppler-broadened resonance
provides a continuous stable operation over several weeks and makes
the setup rather insensitive to mechanical shocks.

At the sensor head the light from the fiber is collimated by a
$f=15\,$mm lens and its polarization is made circular by a polarizing
beamsplitter and a quarter-wave plate placed before the cesium cell.
The light transmitted through the cell is focused into the return
fiber, which guides it to a photodiode. The photocurrent is amplified
by a low-noise transimpedance amplifier. Placing the laser, the
electronics, and the photodiode far away from the sensor head
eliminates magnetic interference generated by those components on the
magnetometer (a photocurrent of 10\,$\mu\mbox{A}$, e.g., produces a
magnetic field of 200\,pT at a distance of 1\,cm). In the present
setup the oscillating-field coil is fed via a twisted-pair conductor,
which represents an effective antenna by which electromagnetic
signals can be coupled into the magnetic shield. In a future stage of
de\-ve\-lop\-ment it is planned to replace this electric lead by an
opto-coupled system.

Multimode fibers were used for ease of light coupling. We found that
a few loops of 3\,cm radius of curvature in the fiber led to
quasi-depolarization of the initially linearly polarized beam,
thereby suppressing noise contributions from polarization
fluctuations. A rigid fixation of the fibers was found necessary to
reduce power fluctuations of the fiber transmission to a level of
$4\times 10^{-5}$ in 1\,Hz bandwidth.

The studies reported below were performed inside closed cylindrical
shields consisting of three layers of MUMETALL (size of the
innermost shield: length 600\,mm, diameter 300\,mm) that reduces the
influence of ambient magnetic field variations. For the measurement
of the noise spectrum (Sec.~\ref{sec:intrsens}) and the study of the
magnetic field stability (Sec.~\ref{sec:measuredsensitivity}) the
shield was improved by three additional cylinders of CO-NETIC
mounted inside of the MUMETALL shield (innermost diameter 230\,mm).
The longitudinal bias field of 2\,$\mu$T, corresponding to a Cs
Larmor frequency of 7\,kHz, is produced by a solenoid (length
600\,mm, diameter 110\,mm) inside the shield and the 8\,mA current
is provided by a specially designed stable current supply.

\subsection{Resonance linewidth}
\label{subsec:resonancelinewidth}

\begin{figure}
  \centering
  \includegraphics[scale=1]{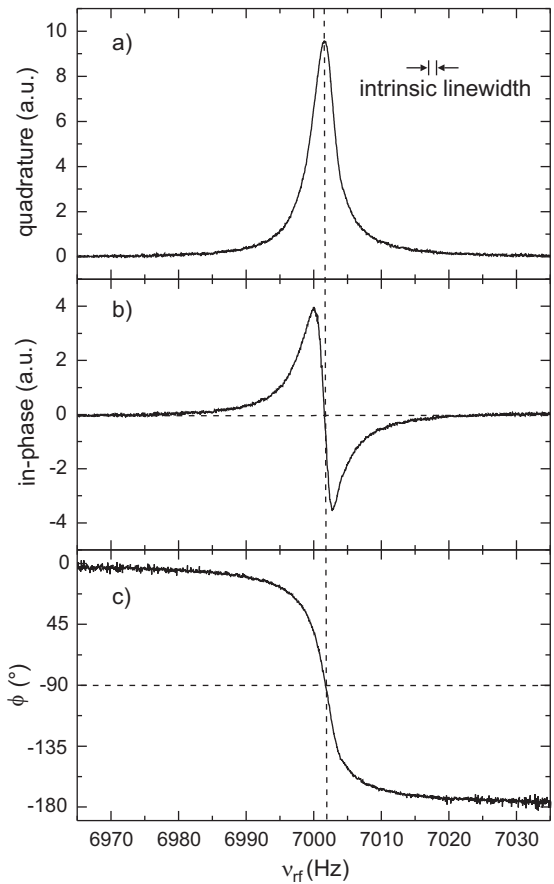}
  \caption{Magnetic resonance spectra obtained by scanning the
  frequency $\nu_{\mathrm{rf}}$ of the oscillating field: a) quadrature component,
  b) in-phase component, c) phase between the oscillating field
  and the modulation of the transmitted power. The Larmor
  frequency $\nu_L$ is 7002.3\,Hz, the power-broadened half linewidth is 2.2\,Hz.
  The intrinsic half linewidth of 1.4\,Hz is indicated.}
  \label{fig:resonance_example}
\end{figure}

\begin{figure}[ht]
\includegraphics{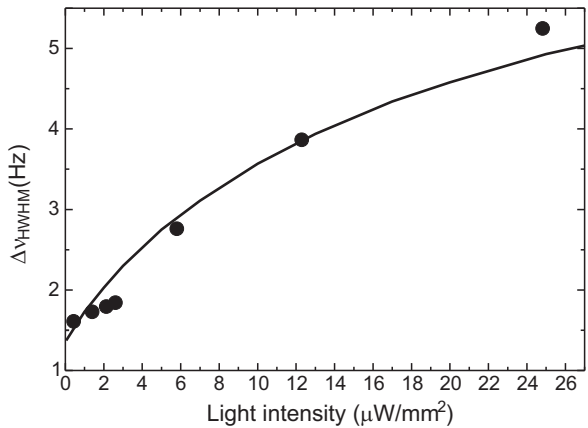}%[scale=0.70]
\caption{Resonance HWHM linewidth as a function of the light
intensity delivered to the cell. The power of the laser beam is given
by $I_L \cdot 2.8\,\mathrm{mm}^2$. The dots represent the widths
obtained from the phase signal of the lock-in amplifier with very low
r.f.~power. The extrapolated intrinsic linewidth is 1.4(1)\,Hz. The
solid line is a one-parameter fit of a numerical calculation to the
data (see text). The size of the symbols represents the vertical
error bars.}\label{fig:intrlinewidth}
%030226
\end{figure}

The lineshapes of the magnetic resonance line are measured with the
magnetometer operating in the open-loop mode
(Fig.~\ref{fig:soplscheme}, mode B). A sinusoidally oscillating
current of frequency $\omega_\mathrm{rf}$ is supplied to the
r.f.~coils by a function generator, whose frequency is ramped across
the Larmor frequency, and the output of the photodiode is
de\-mo\-du\-la\-ted by a lock-in amplifier. Magnetic resonance lines
were recorded for different $B_1$ amplitudes and different values of
the pump light power. Typical resonance lines are shown in
Fig.~\ref{fig:resonance_example}. The lineshapes were fitted by the
function  (\ref{eq:resonance}) to the experimental
$\phi(\nu_{\mathrm{rf}})$ curves, which allows one to infer the
linewidth $\Delta \nu_{\mathrm{HWHM}}$. We recall that the linewidth
is not affected by r.f.~power broadening, but that it is subject to
broadening by the optical pumping process. The dependence of $\Delta
\nu_{\mathrm{HWHM}}$ on the laser intensity
(Fig.~\ref{fig:intrlinewidth}) shows that the optical broadening has
a nonlinear dependence on the light intensity. The minimum or
intrinsic linewidth is determined by extrapolating $\Delta
\nu_{\mathrm{HWHM}}$ to zero light intensity.

For a $J=1/2$ two-level system theory predicts a linear dependence of
the linewidth on the pumping light intensity, as long as stimulated
emission processes from the excited state can be neglected. However,
the magnetic resonance spectrum in the $F=4$ manifold of the Cs
ground state is a superposition of eight degenerate resonances
corresponding to all allowed $\Delta M=\pm 1$ transitions between
adjacent Zeeman levels. The coupling of the $\sigma^+$ polarized
light to the different sublevels depends on their magnetic quantum
number $M_F$ and is given by the corresponding electric dipole
transition matrix elements. As a consequence each of the eight
resonances broadens at a different rate. The observed linewidth
results from the superposition of those individual lines weighted by
the population differences of the levels coupled by the
r.f.~transition and the corresponding magnetic dipole transition
rates. The observed nonlinear dependence of the width on the light
intensity follows from the nonlinear way in which those population
differences and hence the relative weights are changed by the optical
pumping process.

We have calculated the lineshapes of the magnetic resonance lines by
numerically solving the Liouville equation for the ground state
density matrix. Interactions with the optical field as well as the
static and oscillating magnetic fields were taken into account in the
rotating wave approximation. We further assumed an isotropic
relaxation of the spin coherence at a rate given by the
experimentally determined intrinsic linewidth of
Fig.~\ref{fig:intrlinewidth}. The solid curve in that figure
represents the linewidths inferred from the calculated lineshapes.
The calculations used as a variable an optical pumping rate
(proportional to the light power intensity) and the only parameter
used to fit the calculation to the experimental data was the
proportionality constant between the laser intensity and that pump
rate.

The intrinsic linewidth, i.e., the linewidth for vanishing optical
and r.f.~power, is determined by relaxation due to spin exchange
Cs-Cs collisions, Cs-wall collisions, and collisions of the atoms
with the Cs droplet in the reservoir sidearm. The latter contribution
depends on the ratio of the cross section of the constriction in the
sidearm and the inner surface of the spherical cell. With an inner
sidearm diameter of 0.5\,mm that contribution to the HWHM linewidth
can be estimated to be on the order of $\Delta \nu= 1$\,Hz. The
contribution from spin exchange processes at room temperature to the
linewidth can be estimated using the cross-section reported in
\cite{Bev71} to be on the order of 3\,Hz, which is larger than the
measured width. A possible explanation for this discrepancy is the
adsorption of Cs atoms in the paraffin coating \cite{Alex_LIAD},
which may lead to an effective vapor pressure in the cell below its
thermal equilibrium value.

\subsection{Magnetometer mode}
\label{subsec:magnetometer mode}

The actual magnetometry is performed in the phase-stabilized mode
(Fig.~\ref{fig:soplscheme}, mode A) as described above. The
photodiode signal is demodulated by a lock-in amplifier (Stanford
Research Systems, model SR830) locked to the driving r.f.~frequency,
produced by a voltage-controlled oscillator (VCO). The time constant
of the lock-in amplifier was set to $\tau=30\,\mu$s, which
corresponds to a bandwidth of 2.6\,kHz with a $-24$\,dB/octave filter
roll-off. Either the phase (adjusted to be $0^\circ$ on resonance) or
the dispersive in-phase signal of the lock-in amplifier can be used
to control the VCO, and hence to lock its frequency to the center of
the magnetic resonance. Compared to the in-phase signal the phase
signal of the lock-in amplifier has the advantage that the resonance
linewidth is not affected by r.f.~power broadening. However, the
bandwidth of the phase output of the digital lock-in amplifier used
was limited to 200\,Hz by its relatively slow update rate. For the
neutron EDM experiment the magnetometer has to be operated with the
highest possible bandwidth. We therefore chose the in-phase signal
for the following studies. That signal drives the VCO via a feedback
amplifier (integrating and differentiating), which closes the
feedback loop locking the radio frequency to the Larmor frequency.

\section{Performance of the magnetometer}
\label{sec:performance}

\subsection{Magnetometric sensitivity} \label{sec:intrsens}

\begin{figure}
\includegraphics{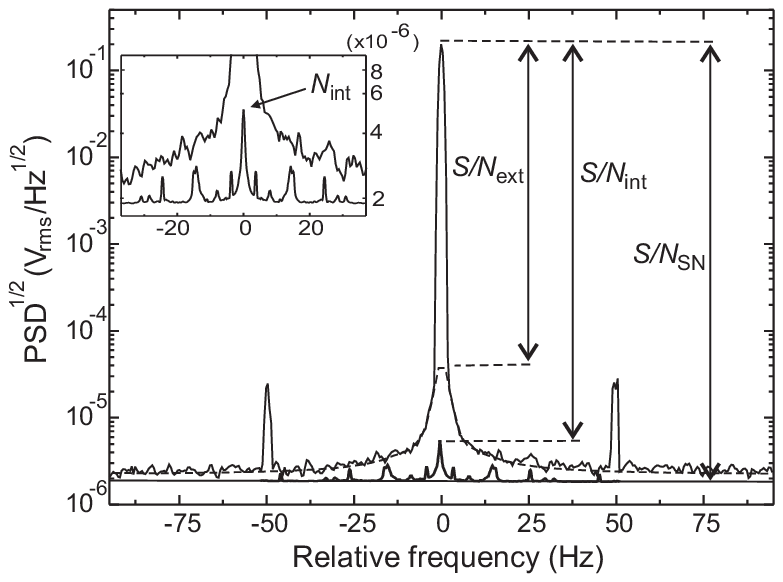}%[scale=.75]
\caption{Square root of the power spectral density (PSD) of the
magnetometer output frequency relative to the Larmor frequency of
$\nu_0=6998$\,Hz (averaged 20 times). The lowest lying curve is the
contribution from laser power noise and $N_\mathrm{int}$ its
contribution at the Larmor frequency (inset). The straight line
indicates the shot noise limit $N_\mathrm{SN}$. The signal-to-noise
ratio $S/N_\mathrm{int}$ is approximately 33000. The sidebands are
due to imperfectly shielded magnetic field components oscillating at
the 50-Hz power line frequency. The signal-to-noise ratio
$S/N_\mathrm{ext}$ due to external field fluctuations is
approximately 4600. All measurements were performed in a 1\,Hz
bandwidth.} \label{fig:FFT}
\end{figure}

We characterize the sensitivity of the magnetometer in terms of the
noise equivalent magnetic flux density (NEM), which is the flux
density change $\delta\!B$ equivalent to the total noise of the
detector signal
\begin{equation}
\delta\!B^2 = \delta\!B_\mathrm{int}^2+\delta\!B_\mathrm{ext}^2~,
\label{eq:sensi}
\end{equation}
with both internal and external contributions:
$\delta\!B_\mathrm{int}$ describes limitations due to noise sources
inherent to the magnetometer proper, while $\delta\!B_\mathrm{ext}$
represents magnetic noise due to external field fluctuations. The
larger of the two contributions determines the smallest flux density
change that the magnetometer can detect. In general the internal NEM
$\delta\!B_\mathrm{int}$ may have several contributions, which may be
expressed as

\begin{equation}
\delta\!B_\mathrm{int}^2 = \left(\frac{1}{\gamma}\times\frac{\Delta
\nu_{\mathrm{HWHM}}} {S/N_\mathrm{SN}}\right)^2+
\sum\limits_{i}{\left(\frac{1}{\gamma}\times\frac{\Delta
\nu_{\mathrm{HWHM}}} {S/N_\mathrm{OPM}^{(i)}}\right)^2}~,
\label{eq:internsens}
\end{equation}
where $S$ is the magnetometer signal, $N_\mathrm{OPM}^{(i)}$ are the
noise levels of the different processes contributing to
$\delta\!B_\mathrm{int}$, and $N_\mathrm{SN}$ the fundamental shot
noise limit of the OPM signal. $\gamma$ is approximately
$3.5\,\mbox{Hz/nT}$ for $^{133}$Cs and $\Delta \nu_{\mathrm{HWHM}}$
is the half width of the resonance (cf. Sec.~\ref{sec:bandwidth}).

The external NEM  $\delta\!B_\mathrm{ext}$ can be parametrized in the
form of Eq.~(\ref{eq:internsens}) with an equivalent signal noise
$N_\mathrm{ext}$ so that Eq.~(\ref{eq:sensi}) can be expressed as
\begin{equation}
\delta\!B = \frac{1}{\gamma}\times\frac{\Delta \nu_{\mathrm{HWHM}}}
{S/N}, \label{eq:sensisimple}
\end{equation}
with $N^2
=N_\mathrm{ext}^2+N_\mathrm{SN}^2+\sum_i{\left(N_\mathrm{OPM}^{(i)}\right)^2}$.

In a strict sense Eqs.~(\ref{eq:internsens}) and
(\ref{eq:sensisimple}) are valid for the open loop operation of the
magnetometer. The parameters $\gamma$ and $\Delta\nu_\mathrm{HWHM}$
do not depend on the mode of operation, whereas $S/N$ may very well
be affected by feedback. If the bandwidth, in which the noise is
measured, is much smaller than the loop bandwidth of 1\,kHz (cf.
Sec.~\ref{sec:bandwidth}), it is reasonable to assume that the power
spectral density, i.e. $S/N$, is not affected by the loop being
opened or closed.

Experimentally the noise levels $N_\alpha$ are determined from a
Fourier ana\-ly\-sis of the photodiode signal, when the magnetometer
is operated in the phase-stabilized mode under optimized parameter
conditions. Each noise level $N$ is defined as the square root of the
integrated (frequency dependent) power spectral density $\rho_S^2$ of
the corresponding signal fluctuations
\begin{eqnarray}
  N&=&\left(\int\limits_{0}^{f_\mathrm{bw}} \rho_S^2
  \mathrm{d}f\right)^{1/2}~,
  \label{eq:deltaBrhoBgeneral}
\end{eqnarray}
where $f_\mathrm{bw}$ is the measurement bandwidth. If the noise is
white or if the bandwidth is much smaller than the width of typical
spectral features in the power spectrum the noise level at a given
frequency $f$ is given by
\begin{eqnarray}
  N=\rho_S \sqrt{f_\mathrm{bw}}= \rho_S /\sqrt{2\tau},
  \label{eq:deltaBrhoBwhite}
\end{eqnarray}
where $\tau$ is the integration time used for calculating the Allan
standard deviation introduced below. Figure~\ref{fig:FFT} shows a
typical Fourier spectrum of the OPM signal. The prominent central
feature corresponds to the Larmor oscillation of the photocurrent at
7\,kHz during the phase-stabilized operation of the OPM. It is the
signal-to-noise ratio at the Larmor frequency which determines the
NEM of the magnetometer. The Larmor peak (carrier) is superposed on
a 20 Hz broad pedestal, which itself lies above a constant noise
floor. The two discrete sidebands originate from magnetic fields
oscillating at the 50\,Hz line frequency.
The pedestal results mainly from imperfectly shielded low-frequency
field fluctuations. The continuous spectrum of such fluctuations is
mixed with the Larmor frequency and therefore appears as a symmetric
background underlying the carrier. We have fitted the pedestal with
a Lorentzian lineshape, from which we infer a signal-to-noise ratio
$S/N_\mathrm{ext}=4600$ at the Larmor frequency, in which
$N_\mathrm{ext}$ corresponds to magnetic field fluctuations
$\delta\!B_\mathrm{ext}$ of 210\,fT in a 1\,Hz bandwidth.

A major contribution to the internal noise $N_\mathrm{int}$ comes
from laser power fluctuations. This can be understood as follows.
The detected photocurrent has the general form
$I_\mathrm{PD}=I_\mathrm{pc}+\Delta I
\cos(\omega_\mathrm{rf}+\phi)$, where $I_\mathrm{pc}$ is the DC
component and $\Delta I$ the amplitude of the AC component, which
depends on the atomic number density and the degree of spin
polarization. If the magnetic resonance signal $\Delta I$ contains
fluctuations at frequency $\omega$, these will appear in
$I_\mathrm{PD}$ at the frequencies $|\omega_\mathrm{rf}\pm \omega|$.
Low frequency power will therefore produce a symmetric background
under the Larmor peak, which also contributes to the pedestal
discussed above. In order to distinguish it from the
$\delta\!B_\mathrm{ext}$ contributions we have measured the low
frequency noise spectrum of the laser power. It shows a 1/$f$-like
behavior for small frequencies, which levels off at the shot noise
value. After mixing with the Larmor frequency its contribution to
the magnetometer noise is shown as the lowest lying curve in
Fig.~\ref{fig:FFT}. The central portion of this spectrum is shown on
an expanded scale in the inset. The sidebands are probably due to
mechanical vibrations, while the central feature extends $\pm
1.5\,\mbox{Hz}$. $N_\mathrm{int}$ represents the contribution of
that power noise to the magnetometer noise at the Larmor frequency.
The direct contribution to $\delta\!B$ from light power noise is
thus negligible compared to the contributions from field
fluctuations and confirms the assumption made above that the latter
dominate the pedestal. The signal-to-noise ratio $S/N_\mathrm{int}$
is 33000 and yields a NEM $\delta\!B_{\mathrm{int}}=29\,\mbox{fT}$
in a 1\,Hz bandwidth, determined by laser power fluctuations.

Far away from the Larmor frequency the measured constant noise floor
exceeds the calculated shot noise level ($N_\mathrm{SN}$) by a
factor of 1.5. This may originate from additional noise sources
related, e.g., to the laser frequency stabilization. The fundamental
limit of the magnetometric sensitivity is determined by the white
shot noise $N_\mathrm{SN}=\sqrt{2 e I_\mathrm{pc} f_\mathrm{bw}}$ of
the DC component of the photocurrent, $I_\mathrm{pc}$, which defines
the ultimate shot noise limited NEM $\delta\!B_\mathrm{SN}$. Under
optimized conditions the photocurrent is 5\,$\mu$A, which yields a
shot noise limited NEM of $\delta\!B_\mathrm{SN} = 10\,\mbox{fT}$ in
a bandwidth of 1\,Hz.

Light shift fluctuations are an additional source of noise. Any
fluctuations of the parameters causing a light shift (laser power
and/or laser frequency detuning) will produce magnetic field
equivalent noise. We will show later that for a 1\,Hz detection
bandwidth this effect gives a negligible contribution to the Fourier
spectrum.

As the internal noise level $\delta\!B_\mathrm{int}$ is much smaller
than the external field fluctuations $\delta\!B_\mathrm{ext}$ the
magnetometer is well suited to measure the characteristics of such
field fluctuations (cf. Sec.~\ref{sec:measuredsensitivity}) and/or to
compensate them using an active feedback loop. The accuracy of such
measurements or the performance of such a stabilization is ultimately
limited by the internal noise of the magnetometer, which under ideal
conditions can reach the shot noise limit.

\subsection{Magnetometer optimization and response bandwidth}
\label{sec:bandwidth}

\begin{figure}
  \includegraphics[scale=1]{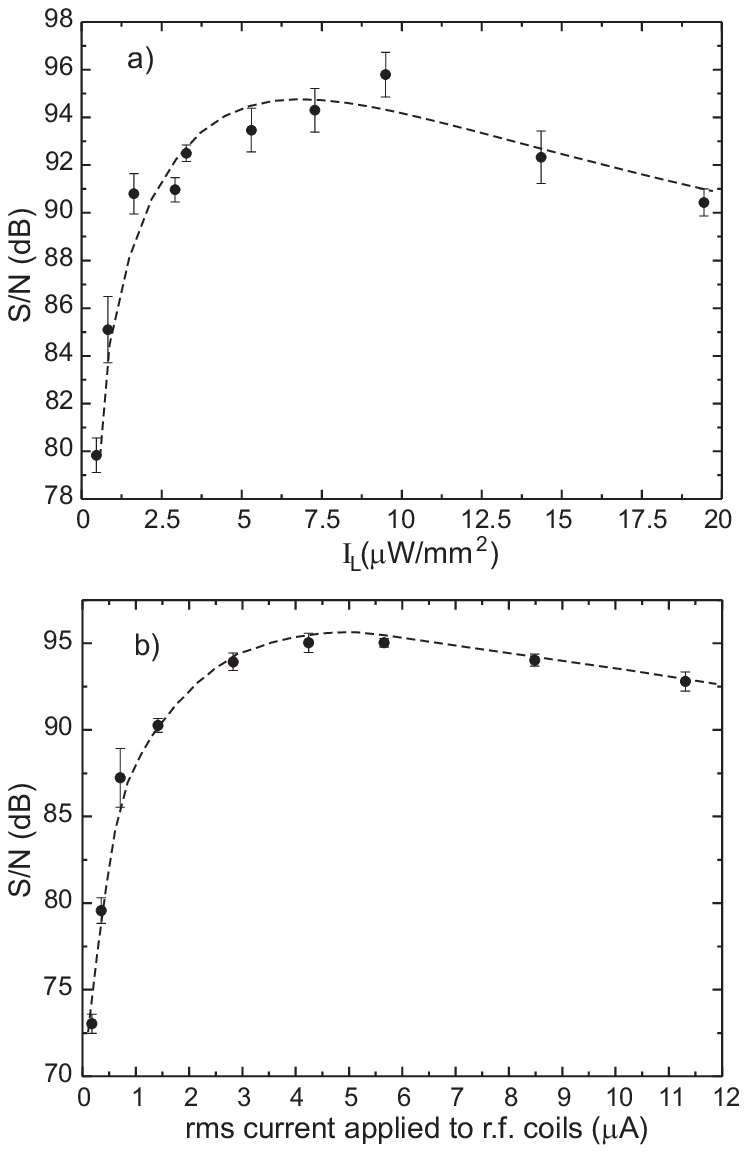}\\
  \caption{Dependence of the experimental signal-to-noise ratio (measured in a 1\,Hz bandwidth) on the light intensity and
  the current applied to the r.f.~coils. The noise was measured 70\,Hz away from the carrier. The r.f.~current in a) was
  8\,$\mu$A$_{\mathrm{pp}}$,
  the light intensity in b) was 7\,$\mu\mathrm{W}/\mathrm{mm}^2$. The dashed lines are drawn to guide the eyes.
  These are typical recordings used to optimize the system parameters.}
  \label{fig:StoN_ILRF}
\end{figure}
\begin{figure}[ht]
  \includegraphics[scale=1]{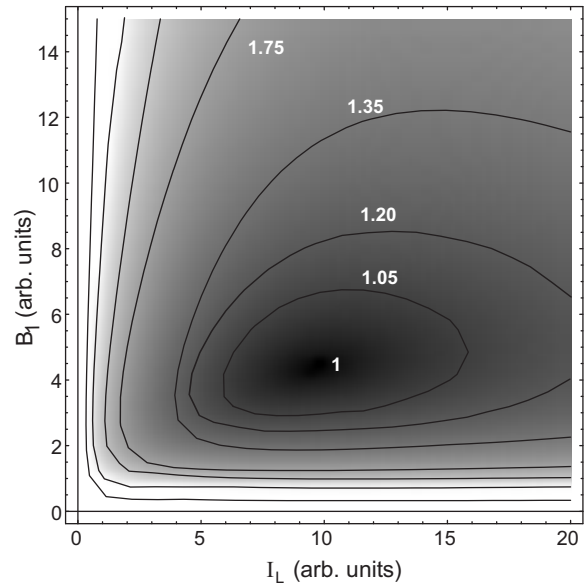}
  \caption{Density plot (in arbitrary units) of the calculated NEM $\delta\!B_\mathrm{SN}$,
   in dependence on the laser intensity $I_L$ and the r.f. amplitude $B_1$.}
  \label{fig:QF_bw}
\end{figure}

According to Eq.~(\ref{eq:internsens}) the sensitivity of the
magnetometer depends on the resonance linewidth $\Delta
\nu_{\mathrm{HWHM}}$ and on the signal-to-noise ratio. For given
properties of the sensor medium (cesium vapor pressure and cell
size) these two properties depend on the two main system parameters,
viz., the laser intensity $I_L$ (or power $P_L$) and the amplitude
$B_1$ of the r.f.\ field. For the application in the neutron EDM
experiment the sensor size and vapor pressure are dictated by the
experimental constraints (fixed geometry and operation at room
temperature), so that the experimental optimization of the
magnetometric sensitivity is performed in the $\left( I_L,\, B_1
\right)$ space by an iterative procedure. Fig.~\ref{fig:StoN_ILRF}
shows examples of signal-to-noise ratio recordings during such an
iteration. The optimum operating point was found for a laser
intensity $I_L$ of $9\,\mu\mbox{W/mm}^2$ and a r.f.\ field amplitude
$B_1$ of $2.7\,\mbox{nT}$. The resonance linewidth under optimum
conditions is $\Delta \nu_{\mathrm{HWHM}}=3.4(1)\,\mbox{Hz}$, which
exceeds the intrinsic linewidth by a factor of 2.4.

In order to investigate the dependence of the NEM on the two
optimization parameters we have calculated that dependence using the
density matrix formalism by assuming that the signal noise is
determined by the shot noise of the photocurrent. The result is shown
in Fig.~\ref{fig:QF_bw} as a density plot. One recognizes a broad
global minimum which is rather insensitive to the parameter values as
it rises only by 5\% when the optimum light and r.f. power are varied
by 50\%.

\begin{figure}
\includegraphics{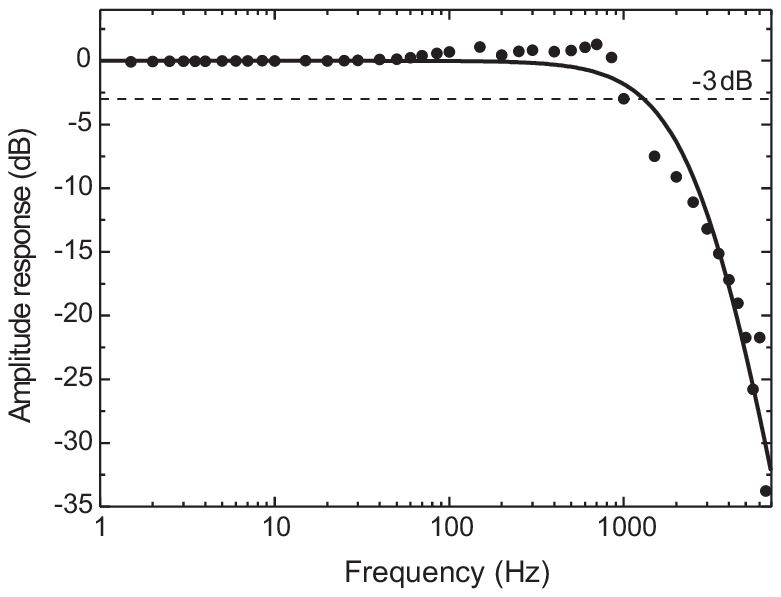}%[scale=0.7]
\caption{Frequency dependence of the magnetometer response to a small
amplitude sinusoidal modulation of the static field $B_0$ (circles).
The solid line indicates the amplitude response of a
4$^\mathrm{th}$-order low-pass filter (-24\,dB/octave roll-off).}
\label{fig:bandwidth}
%030626
\end{figure}

The bandwidth of the magnetometer, i.e., its temporal response to
field changes was measured in the following way: a sinusoidal
modulation of the static magnetic field with an amplitude of 5\,nT
was applied by an additional single wire loop (110\,mm diameter)
wound around the Cs cell. The response of the magnetometer to that
perturbation was measured directly on the VCO input voltage in the
phase-stabilized mode. The result is shown in
Fig.~\ref{fig:bandwidth}. The overall magnetometer response follows
the behavior of a low-pass filter ($-24$\,dB/octave roll-off) with a
-3\,dB point at approximately 1\,kHz. The lock-in time constant was
$30\,\mu\mbox{s}$ which corresponds to a bandwidth of 2.6\,kHz. The
difference is due to additional filters in the feedback loop.

\subsection{Application: Field fluctuations in a magnetic shield}
\label{sec:measuredsensitivity}

\begin{figure}
\includegraphics{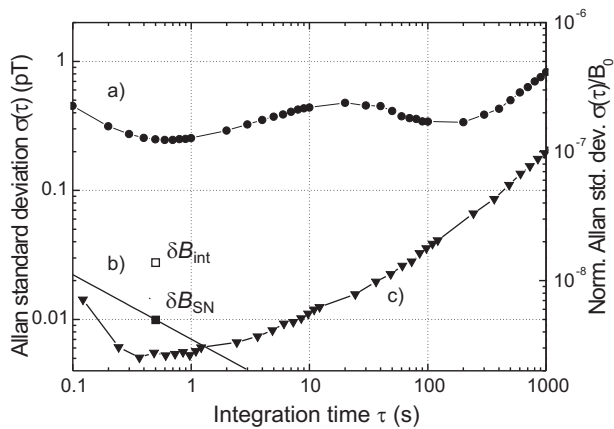}
\caption{a) Allan standard deviation $\delta\!B_\mathrm{ext}$ of the
magnetic flux density inside the magnetic shield ($\bullet$). b) NEM
$\delta\!B_\mathrm{int}$ $(\square)$ limited by laser power
fluctuations; extrapolated NEM $\delta\!B_\mathrm{SN}~(\blacksquare)$
for shot noise limited operation. The slope represents Eq.
(\ref{eq:deltaBrhoBwhite}) assuming a white noise behavior. c)
Measured contributions to $\delta\!B_\mathrm{int}$ from light power
fluctuations ($\blacktriangledown$) with present setup. Solid lines
in a) and c) are drawn to guide the eye. The dwell time of the
frequency counter was 100\,ms. } \label{fig:allan_pl_so}
\end{figure}

External field fluctuations are the dominant contribution to the
noise of the LsOPM when it is operated in the six-layer magnetic
shield. We have used the magnetometer to study the temporal
characteristics of the residual field variations. The Allan standard
deviation \cite{barnes} is the most convenient measure for that
characterization. With respect to the experimental specifications of
the neutron EDM experiment our particular interest is the field
stability for integration times in the range of 100 to 1000\,s. For
that purpose we recorded the Larmor frequency in multiple time
series of several hours with a sampling rate of 0.1\,s by feeding
the photodiode signal, filtered by a resonant amplifier (band-pass
of 200\,Hz width centered at 7\,kHz), to a frequency counter
(Stanford Research Systems, model SR620). From each time series the
Allan standard deviation of the flux density inside the shield was
calculated. A typical result is shown in Fig.~\ref{fig:allan_pl_so}
with both absolute and relative scales. For integration times below
one second the observed fluctuations (curve a) decrease as
$\tau^{-1/2}$, indicating the presence of white field-amplitude
noise. It is characterized by a spectral density of
245\,fT/$\sqrt{\mbox{Hz}}$. Although the Allan standard deviation
represents a different property than the Fourier noise spectrum it
is worthwhile to note that the latter value is comparable with the
NEM $\delta\!B_\mathrm{ext}=210$\,fT of the pedestal in
Fig.~\ref{fig:FFT} discussed above. The field fluctuations reach a
minimum value of approximately 240\,fT for an integration time of
0.7\,s.

The central region of the Allan plot (Fig.~\ref{fig:allan_pl_so}a)
shows a bump for integration times of 1-100 seconds. It is probably
due to fluctuations of the 8\,mA current producing the 2\,$\mu$T
bias field. A magnetic field fluctuation of 200\,fT corresponds to a
relative current stability of $10^{-7}$, i.e., to current
fluctuations of 800\,pA. In an auxiliary experiment we measured the
current fluctuations $\Delta I$ by recording voltage fluctuations
over a series resistor for several hours. We found relative
fluctuations of $\Delta I/I$ in the corresponding Allan plot of the
same order of magnitude as the $\Delta B/B$ fluctuations. It is thus
reasonable to assume that the origin of the plateau in
Fig.~\ref{fig:allan_pl_so}a is due to current fluctuations of the
power supply.

The measurement of the magnetic field during several days shows
fluctuations with a period of one day and an amplitude of about
1\,Hz, superposed by additional uncorrelated drifts. The periodic
fluctuations are probably due to changes of the solenoid geometry
induced by temperature fluctuations. The Allan standard deviations
for integration times exceeding 200\,s are thus determined by
temperature fluctuations and drifts of the laboratory fields, which
are not completely suppressed by the shield.

\subsection{Frequency noise due to light power fluctuations}
\label{sec:lightpowerfluct}

It is well-known that a near-resonant circularly polarized light
field shifts the Zeeman levels in the same way as a static magnetic
field oriented along the light beam. The light shift has
contributions from the AC Stark shift and coherence shift due to
virtual and real transitions \cite{CohTan_TheseTheo}. The AC Stark
shift, and hence the equivalent magnetic field $B_{LS}$ is
proportional to the light intensity $I_L$ and has a dispersive
(Lorentzian) dependence on the detuning of the laser frequency from
the center of the optical absorption line. It is therefore expected
to vanish at the (optical) line center. In our experiment the laser
frequency is locked to the center of a Doppler-broadened hyperfine
component. However, that frequency does not coincide with the
frequency for which the light shift vanishes, because of finite
light shift contributions from the adjacent hyperfine component.
While the two hyperfine components are well separated in the optical
absorption spectra, their corresponding light shift spectra overlap
because of the broad wings of their dispersive lineshapes.

In order to measure the light shift effect we periodically changed
the light power between $P+\Delta P/2$ and $P-\Delta P/2$ and
recorded the corresponding Larmor frequencies. False effects from
drifts of the external magnetic field were suppressed by recording
data over several modulation periods. For each modulation amplitude
$\Delta P$ the Larmor frequency was measured with both $\sigma^+$ and
$\sigma^-$ polarizations by rotating the quarter-wave plate by means
of a mechanical remote control from outside the shield.

\begin{figure}
\includegraphics{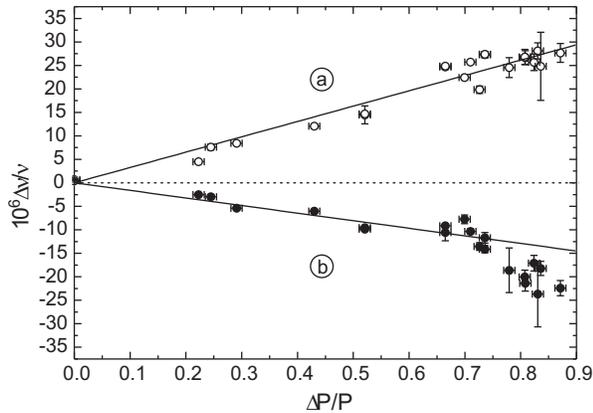}%[scale=0.7]
\caption{Relative light shift of the Larmor frequency as a
function of the relative modulation amplitude $\Delta P$ of the
laser power $P$. Curves (a) and (b) represent measurements with
light of opposite circular polarization.}
\label{fig:Larmorfreqvslightpower}
%031214
\end{figure}

The induced changes of the magnetometer readings for both
polarizations are shown in Fig.~\ref{fig:Larmorfreqvslightpower}. As
anticipated, the shift of the Larmor frequency is proportional to the
modulation amplitude of the light power and changes sign upon
reversing the light helicity. However, it can be seen that the slope
of the light shift depends on the helicity. This asymmetry is the
result of contributions from three distinct effects, which we discuss
only qualitatively here.

(1) The light shift due to virtual transitions (AC Stark shift),
which is proportional to the helicity of the light and thus leads to
a symmetric contribution to the curves of
Fig.~\ref{fig:Larmorfreqvslightpower} (equal in magnitude, but
opposite in sign); (2) the light shift due to real transitions
(coherence shift)\cite{CohTan_TheseTheo}, whose origin is a change
of the effective $g$-factor of the Cs atom due to the fact that with
increasing laser power the atom spends an increasing fraction of its
time in the excited state with a 3 times smaller $g_F$-factor of
opposite sign than that of the ground state; (3) a possible power
dependent change of the capacity of the photodiode and a subsequent
power dependent phase shift of the photocurrent. The latter two
effects yield shifts which have the same sign for both light
polarizations, so that the combined contribution of the three
effects may explain the different magnitudes of the slopes. A
quantitative study of those effects is underway.

Using curve (a) as a worst-case estimate for the fluctuations of the
Larmor frequency due to light power fluctuations we estimated, based
on measured power fluctuations, the resulting magnetic field
fluctuations. The results are shown as triangles in
Fig.~\ref{fig:allan_pl_so}. Light shift fluctuations of the
magnetometer readings are thus one to two orders of magnitude smaller
than residual field fluctuations in the present shield. The light
shift noise can of course be further suppressed by adjusting the
laser frequency to the zero light shift frequency point or better by
actively stabilizing it to that point or by actively stabilizing the
laser power.

\section{Summary and conclusion}
\label{sec:conclusion}

We have described the design and performance of a phase-stabilized
cesium vapor magnetometer. The magnetometer has an intrinsic NEM of
29\,fT, defined as the Allan standard deviation for an bandwidth of
1\,Hz. If the 1/$f$ noise of the laser power can be lowered, e.g.,
by an active power stabilization and the excess white noise floor
can be reduced to the shot noise level the LsOPM should reach a NEM
of 10\,fT for a 1\,Hz bandwidth. The bandwidth of the
phase-stabilized LsOPM is 1\,kHz. We have used the LsOPM to measure
field fluctuations in a six-layer magnetic shield for integration
times between 0.1 and 1000 seconds, whose lowest values were found
to be on the order of 200-300\,fT. Light shift fluctuations, against
which no particular precautions were taken, are one to two orders of
magnitude smaller than the residual field fluctuations in the
shield.

The LsOPM described here compares very favorably with
state-of-the-art lamp-pumped magnetometers. Details on that
comparison will be published elsewhere. It will be a valuable tool
for fundamental physics experiments. The LsOPM presented above meets
the requirements of the neutron-EDM experiment on the relevant time
scales in the range of 100 to 1000~seconds.

\section*{Acknowledgement}

We are indebted to E.~B.~Alexandrov, A.~S.~Pazgalev, and G.~Bison
for numerous fruitful discussions. We acknowledge financial
support from Schweizerischer Nationalfonds, Deutsche
Forschungsgemeinschaft, INTAS and Paul-Scherrer-Institute (PSI).
We thank PSI for the loan of the high-stability current source.

\end{document}